\documentclass[aps,prl,draft,showpacs,twocolumn]{revtex4}

\newcommand{\tr}{\mbox{Tr} }
\newcommand{\ket}[1]{\left | #1 \right \rangle}
\newcommand{\bra}[1]{\left \langle #1 \right |}
\newcommand{\proj}[1]{\ket{#1}\!\!\bra{#1}}

\begin{document}

\title{The Entanglement of Superpositions}

\author{Noah Linden}
\email{n.linden@bristol.ac.uk}
\affiliation{Department of Mathematics, University of Bristol,
Bristol BS8 1TW, United Kingdom}

\author{Sandu Popescu}
\email{s.popescu@bristol.ac.uk}
\affiliation{H H Wills Physics Laboratory, University of
Bristol, Tyndall Avenue, Bristol, BS8 1TL, United Kingdom\\
and
Hewlett-Packard Laboratories, Stoke Gifford, Bristol BS12 6QZ,
United Kingdom}

\author{John A. Smolin}
\email{smolin@watson.ibm.com}
\affiliation{IBM T. J. Watson Research Center, Yorktown Heights,
NY 10598, USA}

\date{5th July 2005}

\begin{abstract}
Given a bipartite quantum state (in arbitrary dimension) and a
decomposition of it as a superposition of two others, we find bounds
on the entanglement of the superposition state in terms of the
entanglement of the states being superposed.  In the case that the two
states being superposed are bi-orthogonal, the answer is simple, and,
for example, the entanglement of the superposition cannot be more than
one e-bit more than the average of the entanglement of the two states
being superposed.  However for more general states, the situation is
very different.
\end{abstract}


\maketitle

\section{Introduction}
The problem we raise in this paper is the following:
Given a state $|\Gamma\rangle$ of two parties, $A$ and $B$, and given a
certain decomposition of it as a superposition of two terms
\begin{equation}
|\Gamma\rangle=\alpha |\Psi\rangle+\beta |\Phi\rangle \ ,
\end{equation}
what is the relation between the entanglement of $|\Gamma\rangle$
and those of the two terms in the superposition?  Given how
central entanglement is to quantum mechanics, and how central
superposition is to entanglement, this question seems to be a
basic one; as far as we are aware, however, little is known about
it.  This is particularly surprising for bipartite pure states, as
for them at least the measure of entanglement is completely
understood--the entanglement of a bipartite pure state is the von
Neumann entropy of the reduced state of either of the parties
\cite{BBPS}:
\begin{equation}
E(\Psi)\equiv S\big(\tr_A\!\proj{\Psi}\big)=S\big(\tr_B\!\proj{\Psi}\big)
\end{equation}

Before embarking on our study, it is worth making some observations.
To start with, at first sight it seems unlikely that
there could be any relation at all. Indeed, entanglement is a
global property of a state, originating precisely from the
superposition of different terms; looking at each term separately
seems  completely to miss the point. For example, consider a state
of two qubits
\begin{equation}|\gamma\rangle={1\over{\sqrt2}}|0\rangle|0\rangle+{1\over{\sqrt2}}|1\rangle|1\rangle.\end{equation}
Each term by itself is unentangled, yet their superposition is a
maximally entangled state of the qubits. On the other hand,
consider
\begin{equation}|\gamma'\rangle={1\over{\sqrt2}}|\Phi^+\rangle+{1\over{\sqrt2}}|\Phi^-\rangle\end{equation}
 where
\begin{equation}|\Phi^\pm\rangle={1\over{\sqrt2}}|0\rangle|0\rangle\pm{1\over{\sqrt2}}|1\rangle|1\rangle\end{equation}
Each of the terms in the superposition is maximally entangled, yet the
superposition itself is unentangled.

We also note that in general, two states of high fidelity to one another--i.e. {\em they are almost the
same state}--do not necessarily have nearly the same entanglement.  That is
when $|\langle\psi|\phi\rangle|^2\rightarrow 1$ in general it is not true that
$E(\psi)\rightarrow E(\phi)$.

For example let
\begin{eqnarray}
\label{highfi}\ket{\phi}&=&\ket{0}\!\ket{0}\ {\rm and}\\
\ket{\psi}&=&{\sqrt{1-\epsilon}}\ket{\phi} + \sqrt{\frac{\epsilon}{d}}
\Big[|1\rangle|1\rangle+|2\rangle|2\rangle\ldots+|d\rangle|d\rangle\Big] \ .\nonumber
\end{eqnarray}
In this case $E(\phi)=0$ but
\begin{equation}E(\psi)= - (1-\epsilon)\log_2
(1-\epsilon)-d\left( {{\epsilon}\over d}\log_2{{\epsilon}\over d}\right)\approx \epsilon\log_2 d\ .
\end{equation}
The fidelity $|\langle\psi|\phi\rangle|^2=1-\epsilon$ approaches
one for small $\epsilon$, but for any $\epsilon$ we can pick a $d$
such that the difference in the entanglements of $\ket\phi$ and
$\ket\psi$ is however large we like.  The amount the entanglement
of two states of fixed dimension can differ as a function of
fidelity is bounded using Fannes's inequality \cite{fannes}.  In
infinite dimensions no such bound applies and entanglement is not
a continuous function.

On the other hand, suppose that we have a state with large number
of Schmidt terms in its decomposition. It is obvious that by
adding a small number of supplementary terms with small overall
weight (and then normalize the resulting state) one cannot affect
the overall entanglement too much. This leads us to think that
despite the previous arguments, there is a relation between the
entanglement of a state and the individual terms that by
superposition yield the state.

\section{Bi-orthogonal states}
The simplest case is when the two states we are superposing,
$\ket{\Phi_1}$ and $\ket{\Psi_1}$ are bi-orthogonal, i.e.
\begin{eqnarray}
\tr_A\Big(\tr_B(\proj{\Phi_1})\tr_B(\proj{\Psi_1})\Big)
&=&\nonumber\\
\tr_B\Big(\tr_A(\proj{\Phi_1})\tr_A(\proj{\Psi_1})\Big)&=&0.
\end{eqnarray}
Up to local unitary transformations,
\begin{eqnarray}
\ket{\Phi_1}&=&\sum_{i=1}^{d_1}a_i\ket i\!\ket i
\nonumber\\
\ket{\Psi_1}&=&\sum_{i=d_1+1}^{d}b_i\ket i\!\ket i,
\end{eqnarray}
where $a_i$ and $b_i$ are positive and real.  Since Alice's reduced states for $\ket{\Phi_1}$ and $\ket{\Psi_1}$ are diagonal
in the same basis, is not difficult
to calculate directly that the entanglement of the superposition
$\ket{\Gamma_1}=\alpha\ket{\Phi_1}+\beta\ket{\Psi_1}$ is given
by
\begin{eqnarray}
E(\Gamma_1) =|\alpha|^2 E({\Phi_1}) + |\beta|^2 E({\Psi_1}) + h_2(|\alpha|^2),\label{bi-orthog-equality}
\end{eqnarray}
where $h_2(x)=-x\log_2 x -(1-x)\log_2(1-x)$ is the binary entropy function, and we take $|\alpha|^2 + |\beta|^2 = 1$.

In fact the following inequalities hold for any density matrices \cite{entropy-ineqs}; these will
be used repeatedly in what follows ($S(\rho)$ denotes the von Neumann entropy of $\rho$):
\begin{equation}
|\alpha|^2 S({\rho}) + |\beta|^2 S(\sigma) \leq S(|\alpha|^2 \rho + |\beta|^2 \sigma)\label{ineq1}
\end{equation}
and
\begin{equation}
S(|\alpha|^2 \rho + |\beta|^2 \sigma)\leq |\alpha|^2 S({\rho}) + |\beta|^2 S(\sigma)+ h_2(|\alpha|^2).\label{ineq2}
\end{equation}
There is equality in (\ref{ineq2}) if and only if $\rho$ and $\sigma$ are orthogonal.
Since $\ket{\Phi_1}$ and $\ket{\Psi_1}$ are bi-orthogonal,
their reduced density matrices are orthogonal so we could have used (\ref{ineq2})
rather than direct calculation to give (\ref{bi-orthog-equality}).

Let us use the following notation for the expression on the right-hand-side of equation (\ref{bi-orthog-equality}):
\begin{eqnarray}
\Upsilon(\Phi,\Psi,\alpha)\equiv |\alpha|^2 E({\Phi}) + |\beta|^2 E({\Psi}) + h_2(|\alpha|^2).
\end{eqnarray}
Thus for bi-orthogonal states, the ratio
\begin{equation}
{{E(\Gamma_1)} \over {\Upsilon(\Phi_1,\Psi_1,\alpha)}}  = 1\ .\label{bi-orthog-equality2}
\end{equation}
In addition the maximum increase of entanglement is bounded:
\begin{eqnarray}
E(\Gamma_1) -\Big(|\alpha|^2 E({\Phi_1}) + |\beta|^2 E({\Psi_1}) \Big)\leq 1,
\label{increasebound}
\end{eqnarray}
independent of the dimension.

We also point out that if we {\em mix} rather than superpose two pure
states, the entanglement of formation \cite{purification,bdsw} is at most the
average of the entanglement of the individual states.

However we will soon see that any intuition we might have gained by
considering the case of bi-orthogonal states is misleading.

\section{Orthogonal (but not necessarily bi-orthogonal) states}
We now prove the following result. Given two states $\ket{\Phi_2}$ and $\ket{\Psi_2}$
which are orthogonal but not
necessarily bi-orthogonal, the entanglement of the superposition
\begin{eqnarray}
\ket{\Gamma_2}=\alpha\ket{\Phi_2} + \beta\ket{\Psi_2},
\end{eqnarray}
(where $|\alpha|^2 +
|\beta|^2 = 1$, so that $\ket{\Gamma_2}$ is normalized) satisfies
\begin{eqnarray}
& &E(\alpha{\Phi_2} + \beta{\Psi_2})\nonumber\\
& &\quad\leq 2\Big( |\alpha|^2 E({\Phi_1}) + |\beta|^2 E({\Psi_1}) + h_2(|\alpha|^2)\Big).\label{orthog-inequality}
\end{eqnarray}

To prove this,
consider that Alice, in addition to Hilbert
space $\mathcal{H}_A$, has a qubit with Hilbert space denoted
$\mathcal{H}_a$.  And consider the state
\begin{eqnarray}
\ket{\Delta_2}=\alpha{\ket 0}_a \ket{\Phi_2}_{AB} + \beta{\ket 1}_a\ket{\Psi_2}_{AB}.
\label{Delta2}
\end{eqnarray}
Bob's reduced state for $\ket{\Delta_2}$ is
\begin{eqnarray}
\rho_B=|\alpha|^2\tr_A\Big(\proj{\Phi_2}\Big)+|\beta|^2\tr_A\Big(\proj{\Psi_2}\Big).
\end{eqnarray}
The inequality (\ref{ineq2}) shows that
\begin{eqnarray}
S(\rho_B)&\leq& |\alpha|^2 S\big(\tr_A\!\proj{\Phi_2}\big) \\
& &+\, |\beta|^2 S\big(\tr_A\!\proj{\Psi_2}\big) + h_2(|\alpha|^2).\nonumber
\end{eqnarray}
However $\rho_B$ may also be written
\begin{eqnarray}
\rho_B&=&{1\over 2}\tr_A\Big[\Big(\alpha\ket{\Phi_2}+\beta\ket{\Psi_2}\Big)\Big(\bar\alpha\bra{\Phi_2}+\bar\beta\bra{\Psi_2}\Big)\Big]\\
& &\quad+{1\over 2}\tr_A\Big[\Big(\alpha\ket{\Phi_2}-\beta\ket{\Psi_2}\Big)\Big(\bar\alpha\bra{\Phi_2}-\bar\beta\bra{\Psi_2}\Big)\Big].\nonumber
\end{eqnarray}
Thus (\ref{ineq1}), shows that
\begin{eqnarray}
& &{1\over 2}S\Big(\tr_A\Big[\Big(\alpha\ket{\Phi_2}+\beta\ket{\Psi_2}\Big)\Big(\bar\alpha\bra{\Phi_2}+\bar\beta\bra{\Psi_2}\Big)\Big]\Big)\nonumber\\
& &\quad+{1\over 2}S\Big(\tr_A\Big[\Big(\alpha\ket{\Phi_2}-\beta\ket{\Psi_2}\Big)\Big(\bar\alpha\bra{\Phi_2}-\bar\beta\bra{\Psi_2}\Big)\Big]\Big)\nonumber\\
& &\quad\leq S(\rho_B).
\end{eqnarray}
Thus
\begin{eqnarray}
& &{1\over 2}E(\alpha{\Phi_2} + \beta{\Psi_2}) + {1\over 2}E(\alpha{\Phi_2} - \beta{\Psi_2})\nonumber\\
& &\quad\leq \Big( |\alpha|^2 E({\Phi_1}) + |\beta|^2 E({\Psi_1}) + h_2(|\alpha|^2)\Big).
\end{eqnarray}
Since $E(\alpha{\Phi_2} - \beta{\Psi_2})\geq 0$, we deduce the advertised inequality (\ref{orthog-inequality}).

The inequality may also be written
\begin{equation}
{E(\alpha{\Phi_2}+\beta{\Psi_2})\over
\Upsilon(\Phi_2,\Psi_2,\alpha)}\leq 2.
\label{ratiobound}
\end{equation}
One may wonder whether the factor of two on the right-hand-side of this equation is an artifact of our proof, and whether in fact
the factor should be one as in (\ref{bi-orthog-equality2}).  As we now show, even for qubits, one can get as close as we wish
to the ratio two in this equation.  For consider the following choices:
\begin{eqnarray}
\ket{\phi_2}&=& \ket 0\!\ket 0\nonumber\\
\ket{\psi_2} &=& \sqrt{y/2}\ket 0\!\ket 1 +\sqrt{y/2}\ket 1\!\ket 0 - \sqrt{1-y}
\ket 1\!\ket 1\nonumber\\
\alpha&=&xy;\quad \beta=\sqrt{1-\alpha^2}
\end{eqnarray}
where $x$ and $y$ are real parameters.  We are interested in the behavior of this family of states
as $y$ tends to zero, with $x$ fixed.   $\ket{\phi_2}$ is unentangled, and as $y$ tends to zero (with $x$ fixed),  $\ket{\psi_2}$ and
$\ket{\gamma_2}=\alpha\ket{\phi_2} + \beta\ket{\psi_2}$ both get closer and closer to being unentangled.
It is not difficult to check that
\begin{equation}
\lim_{y\rightarrow 0}{E(\alpha{\phi_2}+\beta{\psi_2})\over
\Upsilon(\phi_2,\psi_2,\alpha)}
=
{(1 + 2x)^2\over 1 + 4x^2}.
\end{equation}
We note that this limit is $2$ for $x=1/2$.

Since the states in this case are close to being unentangled, the
example might seem to be a trick of the limiting behavior and possibly
uninteresting.  It might be thought that one can only achieve equality
in the bound (\ref{ratiobound}) for essentially unentangled states,
and that the {\em increase} in entanglement could never violate the bound
(\ref{increasebound}).
However, in larger dimensions than qubits this is not the case.  Consider
the following example when Alice and Bob both have Hilbert spaces of dimension $d$:
\begin{eqnarray}
\ket{\phi_2^\prime}&=& {1\over\sqrt{2}}\Big(\ket 1\!\ket 1+{1\over\sqrt{d-1}}\Big[\ket 2\!\ket 2 + \ket 3\!\ket 3\ldots\ket d\!\ket d\Big]\Big)\nonumber\\
\!\ket{\psi_2^\prime}&=& {1\over\sqrt{2}}\Big(\ket 1\!\ket 1-{1\over\sqrt{d-1}}\Big[\ket 2\!\ket 2 + \ket 3\!\ket 3\ldots\ket d\!\ket d\Big]\Big)\nonumber\\
\alpha&=&-\beta={1\over\sqrt{2}}
\end{eqnarray}
The entanglement of $\ket{\phi_2^\prime}$ and $\ket{\psi_2^\prime}$ is ${1\over 2}\log_2(d-1) + 1 $; and the entanglement of
$\alpha\ket{\phi_2^\prime}+\beta\ket{\psi_2^\prime}$ is $\log_2(d-1)$.  Thus
\begin{equation}
{E(\alpha{\phi_2^\prime}+\beta{\psi_2^\prime})\over
\Upsilon(\phi_2^\prime,\psi_2^\prime,\alpha)}\rightarrow 2,
\end{equation}
as $d\rightarrow \infty$ and
the increase in entanglement is
\begin{eqnarray}
E(\alpha{\phi_2^\prime}+\beta{\psi_2^\prime})&-&\Big( |\alpha|^2 E({\phi_2^\prime}) + |\beta|^2 E({\psi_2^\prime}) \Big)\nonumber\\
& &\quad ={1\over 2}\log_2(d-1) - 1.
\end{eqnarray}
Thus the increase in entanglement can be greater than one e-bit, and
in fact unbounded.  For this example the increase in entanglement is
only greater than one e-bit for $d>17$.  However, using numerical
searches, we have found examples even for $d=3$ for which the increase
in entanglement is more than one e-bit.

\section{Arbitrary states}
The most general case, when the two states we are superposing are non-orthogonal, is also interesting.
In this case we may prove the following inequality:  Let $\ket{\Phi_3}$ and $\ket{\Psi_3}$ be
 normalized but otherwise arbitrary, and as before we take $|\alpha|^2 + |\beta|^2 = 1$.  Then
 \begin{eqnarray}
& &\|\alpha\ket{\Phi_3}-\beta\ket{\Psi_3}\|^2 E(\alpha{\Phi_3} + \beta{\Psi_3})\leq\nonumber\\
& &\quad{2}\Big( |\alpha|^2 E({\Phi_3}) + |\beta|^2 E({\Psi_3}) + h_2(|\alpha|^2)\Big)\label{general-ineq}
\end{eqnarray}
The notation $E(\alpha{\Phi_3}+\beta{\Psi_3})$ denotes the entanglement of the normalized version of the state
$\alpha\ket{\Phi_3}+\beta\ket{\Psi_3}$.

To prove (\ref{general-ineq}) again let us consider an expression of the form (\ref{Delta2})
\begin{eqnarray}
\ket{\Delta_3}=\alpha{\ket 0}_a \ket{\Phi_3}_{AB} + \beta{\ket 1}_a\ket{\Psi_3}_{AB}.\label{Delta3}
\end{eqnarray}
Although $\ket{\Delta_3}$  is normalized, the state
\begin{eqnarray}
\ket{\Gamma_3}=\alpha\ket{\Phi_3} + \beta\ket{\Psi_3}
\end{eqnarray}
need not be.

Now, as before,  Bob's reduced state for $\ket{\Delta_3}$ can be written in two ways:
\begin{eqnarray}
\rho_B=|\alpha|^2\tr_A\Big(\proj{\Phi_3}\Big)+|\beta|^2\tr_A\Big(\proj{\Psi_3}\Big)
\end{eqnarray}
and
\begin{widetext}
\begin{eqnarray}
\rho_B&=&
\quad{\|\alpha\ket{\Phi_3}+\beta\ket{\Psi_3}\|^2\over 2}\tr_A\Big[\Big(
{\alpha\ket{\Phi_3}+\beta\ket{\Psi_3}\over
\|\alpha\ket{\Phi_3}+\beta\ket{\Psi_3}\|}
\Big)
\Big(
{\bar\alpha\bra{\Phi_3}+\bar\beta\bra{\Psi_3}
\over
\|\alpha\ket{\Phi_3}+\beta\ket{\Psi_3}\|}
\Big)\Big]\nonumber\\
& &\quad+\,{\|\alpha\ket{\Phi_3}-\beta\ket{\Psi_3}\|^2\over 2}\tr_A\Big[\Big(
{\alpha\ket{\Phi_3}-\beta\ket{\Psi_3}\over
\|\alpha\ket{\Phi_3}-\beta\ket{\Psi_3}\|}
\Big)
\Big(
{\bar\alpha\bra{\Phi_3}-\bar\beta\bra{\Psi_3}
\over
\|\alpha\ket{\Phi_3}-\beta\ket{\Psi_3}\|}
\Big)\Big].\nonumber
\end{eqnarray}
\end{widetext}
We have explicitly written $\rho_B$ as a mixture of trace one operators.
So now using (\ref{ineq1}) and (\ref{ineq2}) we deduce that
\vspace{10pt}
\begin{eqnarray}
& &\|\alpha\ket{\Phi_3}+\beta\ket{\Psi_3}\|^2 E(\alpha{\Phi_3} + \beta{\Psi_3})\leq\\
& &\quad{2}\Big( |\alpha|^2 E({\Phi_3}) + |\beta|^2 E({\Psi_3}) + h_2(|\alpha|^2)\Big),\nonumber
\end{eqnarray}
or equivalently
\begin{equation}
{E(\alpha{\Phi_3}+\beta{\Psi_3})\over
\Upsilon(\Phi_3,\Psi_3,\alpha)}\leq {2\over\|\alpha\ket{\Phi_3}+\beta\ket{\Psi_3}\|^2}.
\end{equation}

We do not know whether this bound is the best possible; we suspect not.  However, unlike the case
where the two superposed states are orthogonal,  for which the ratio
\begin{equation}
{E(\alpha{\Phi_3}+\beta{\Psi_3})\over
\Upsilon(\Phi_3,\Psi_3,\alpha)}
\end{equation}
is bounded by two, this ratio is unbounded for non-orthogonal states.  For consider
\begin{eqnarray}
\ket{\phi_3}&=& \ket 1\!\ket 1\nonumber\\
\ket{\psi_3}&=& \sqrt{1-\epsilon}\ket 1\!\ket 1-{\epsilon\over\sqrt{d}}\Big[\ket 1\!\ket 1+ \ket 2\!\ket 2 \ldots\ket d\!\ket d\Big]\nonumber\\
\alpha&=&{\sqrt{1-\epsilon}\over\sqrt{2-\epsilon}};\quad    \beta={-1\over\sqrt{2-\epsilon}}.
\end{eqnarray}
In this case
\begin{equation}
{\alpha\ket{\phi_3}+\beta\ket{\psi_3}\over
\|\alpha\ket{\phi_3}+\beta\ket{\psi_3}\|} =
{1\over\sqrt{d}}\Big[\ket 1\!\ket 1+ \ket 2\!\ket 2 \ldots\ket d\!\ket d\Big]
\end{equation}
So $E(\alpha{\phi_3}+\beta{\psi_3})=\log_2 \!d$, and for fixed $d$, we can
let $\epsilon$ be as small as we like so that $E(\psi_3)\approx 0 $ and
$h_2(\alpha^2)\approx 1 $.  Hence as $\epsilon\rightarrow 0$,
\begin{equation}
{E(\alpha{\phi_3}+\beta{\psi_3})\over
\Upsilon(\phi_3,\psi_3,\alpha)}\rightarrow \log_2\!d.
\end{equation}
This example is also interesting since the increase in entanglement
\begin{equation}
E(\alpha{\phi_3}+\beta{\psi_3})-\Big(|\alpha|^2 E({\phi_3}) +
|\beta|^2 E({\psi_3}) \Big)\rightarrow \log_2\!d,
\end{equation}
as $\epsilon\rightarrow 0$, which is the maximum possible increase in dimension $d$.
Notice that the trick used here is quite similar to that used in (\ref{highfi})
which exhibits two states of high fidelity but very different entanglement.  Here,
we take $\epsilon$ still smaller, resulting in two states of high fidelity, nearly
the same entanglement, but vastly different Schmidt ranks.  The large increase of
entanglement comes about when the normalization of the superposition
$\ket{\gamma_3}=\alpha\ket{\phi_3}+\beta\ket{\psi_3}$ supplies weight to
those many Schmidt terms.

We end by noting that the methods we have used yield straightforward generalizations of
our results to cases where there are more than two terms in the superposition.
\begin{acknowledgments}
NL and SP  thank the EU for support through the European Commission
project RESQ (contract IST-2001-37559);  NL, SP and JAS thank the UK EPSRC for
support through the Interdisciplinary Research Collaboration in Quantum Information Processing;
JAS thanks the US National Security Agency and the Advanced Research
and Development Activity for support through contract
DAAD19-01-C-0056.

\end{acknowledgments}

\end{document}